\documentclass[12pt]{article}
 \makeatletter\ifcase\@ptsize\font\teneufm=eufm10
\font\seveneufm=eufm7\font\fiveeufm=eufm5
\font\teneusm=eusm10\font\seveneusm=eusm7
\font\fiveeusm=eusm5\or\font\teneufm=eufm10 scaled
\magstephalf\font\seveneufm=eufm7\font\fiveeufm=eufm5
\font\teneusm=eusm10 scaled\magstephalf
\font\seveneusm=eusm7\font\fiveeusm=eusm5\or \font\teneufm=eufm10
scaled\magstep1\font\seveneufm=eufm7
\font\fiveeufm=eufm5\font\teneusm=eusm10 scaled\magstep1
\font\seveneusm=eusm7\font\fiveeusm=eusm5\fi

\newfam\eufmfam\newfam\eusmfam\textfont\eufmfam=\teneufm
\scriptfont\eufmfam=\seveneufm \scriptscriptfont\eufmfam=\fiveeufm
\textfont\eusmfam=\teneusm\scriptfont\eusmfam=\seveneusm
\scriptscriptfont\eusmfam=\fiveeusm

\def\frak{\ifmmode\let\next\frak@\else
\def\next{\errmessage{Use\string\frak\space only in math
mode}}\fi\next}\def\frak@#1{{\frak@@{#1}}}
\def\frak@@#1{\fam\eufmfam#1}
\def\sh{\ifmmode\let\next\sh@\else
\def\next{\errmessage{Use\string\sh\space only in math
mode}}\fi\next}\def\sh@#1{{\sh@@{#1}}}
\def\sh@@#1{\fam\eusmfam#1}

\ifcase\@ptsize\font\tenmsa=msam10\font\sevenmsa=msam7
\font\fivemsa=msam5\font\tenmsb=msbm10
\font\sevenmsb=msbm7\font\fivemsb=msbm5\or \font\tenmsa=msam10
scaled\magstephalf \font\sevenmsa=msam7\font\fivemsa=msam5
\font\tenmsb=msbm10 scaled\magstephalf
\font\sevenmsb=msbm7\font\fivemsb=msbm5\or \font\tenmsa=msam10
scaled\magstep1\font\sevenmsa=msam7
\font\fivemsa=msam5\font\tenmsb=msbm10 scaled\magstep1
\font\sevenmsb=msbm7\font\fivemsb=msbm5\fi

\newfam\msafam\newfam\msbfam\textfont\msafam=\tenmsa
\scriptfont\msafam=\sevenmsa
\scriptscriptfont\msafam=\fivemsa\textfont\msbfam=\tenmsb
\scriptfont\msbfam=\sevenmsb \scriptscriptfont\msbfam=\fivemsb

\def\Bbb{\ifmmode\let\next\Bbb@\else
\def\next{\errmessage{Use\string\Bbb\space only in math
mode}}\fi\next}\def\Bbb@#1{{\Bbb@@{#1}}}
\def\Bbb@@#1{\fam\msbfam#1}\def\hexnumber@#1{\ifnum#1<10
\number#1\else\ifnum#1=10 A\else\ifnum#1=11 B\else\ifnum#1=12
C\else\ifnum#1=13 D\else\ifnum#1=14 E\else\ifnum#1=15
F\fi\fi\fi\fi\fi\fi\fi}
\def\msa@{\hexnumber@\msafam}\def\msb@{\hexnumber@\msbfam}
\mathchardef\square="0\msa@03

\makeatother

\newcommand {\CMP}  {Comm. Math. Phys.}

\newcommand {\IJMP} {Int. J. Mod. Phys.}
\newcommand {\JHEP} {JHEP}
\newcommand {\MPL}  {Mod. Phys. Lett.}
\newcommand {\NP}   {Nucl. Phys.}

\newcommand {\PL}   {Phys. Lett.}

\newcommand {\PRP}  {Phys. Rep.}

\newcommand{\half}{\frac{1}{2}}

\def\a{\alpha}
\def\l{\lambda}

\def\b{\beta}

\def\G{\Gamma}
\def\d{\delta}

\def\s{\sigma}

\def\t{\theta}

\def\p{\partial}

\def\etab{\bar{\eta}}

\def\bl{\bar\lambda}

\def\Tr{\textrm{Tr}}

\def\be{\begin{equation}}
\def\ee{\end{equation}}
\def\ba{\begin{array}} \def\ea{\end{array}}
\def\bea{\begin{eqnarray}}
\def\eea{\end{eqnarray}}
\def\nn{\nonumber}

\def\ar{\rightarrow}

\begin{document}
\thispagestyle{empty}

\begin{flushright}
DFPD02/TH/14\\
EDO-EP-44\\
hep--th/0206104\\
June, 2002\\
\end{flushright}
\vspace{20pt}

\baselineskip15pt

\begin{center}
{\large\bf The Superembedding Origin of the Berkovits Pure Spinor Covariant
Quantization of Superstrings \vskip 1mm}

\vspace{10mm}

Marco Matone$^{a,b}$, Luca Mazzucato$^a$, Ichiro Oda$^c$, Dmitri
Sorokin$^{a,b,d}$ and Mario Tonin$^{a,b}$

\vspace{10mm}

${}^a$ Dipartimento di Fisica, Universit\`a degli Studi di Padova,\\
Via F. Marzolo 8, 35131 Padova, Italy\\

\vspace{5mm}

${}^b$ Istituto Nazionale di Fisica Nucleare, Sezione di Padova\\

\vspace{5mm}

${}^c$ Edogawa University,
474 Komaki, Nagareyama City, Chiba 270-0198, Japan\\

\vspace{5mm}

${}^d$ Institute for Theoretical Physics, NSC KIPT, 61108 Kharkov,
Ukraine
\end{center}

\vspace{5mm}
\begin{abstract}
\noindent We show that the pure spinor formalism proposed by
Berkovits to covariantly quantize superstrings is a gauge fixed,
twisted version of the complexified $n=2$ superembedding
formulation of the superstring. This provides the Berkovits
approach with a geometrical superdiffeomorphism invariant ground.
As a consequence, the absence of the worldsheet
(super)diffeomorphism ghosts in the pure spinor quantization
prescription and the nature of the Berkovits BRST charge and
antighost are clarified. Since superembedding is classically
equivalent to the Green--Schwarz formulation, we thus also relate
the latter to the pure spinor construction.
\end{abstract}

\newpage

\section{Introduction}

The superembedding is a geometrical description of the dynamics of
superbranes by means of a specific embedding of worldvolume
supersurfaces into target superspaces. It has proved to be a
powerful method for studying various aspects of superbranes at the
classical level (see \cite{report} for a review). For instance,
the superembedding description has explained the geometrical
nature of the $\kappa$--symmetry of the Green--Schwarz formulation
\cite{siegel}, which turns out to be a conventional extended local
supersymmetry of the embedded superworldvolume (as was first
proved for superparticles in \cite{stv,gal}, for superstrings in
\cite{berk}--\cite{igo} and then for all the other superbranes
\cite{pt,bpstv,hs}). An original purpose of the superembedding
approach has been to use the worldvolume supersymmetry nature of
the $\kappa$--symmetry for making progress in solving a
long--standing problem of the covariant quantization of the
Green--Schwarz superstrings, the problem being just caused by the
$\kappa$--symmetry of the latter. To this end doubly (worldsheet +
target space) supersymmetric actions for superstrings were
constructed in which the whole or part of the $\kappa$--symmetry
was traded for a manifest irreducible worldsheet
supersymmetry.\footnote{In earlier versions of doubly
supersymmetric extended objects \cite{ss} the $\kappa$--symmetry
remained an independent symmetry, and these models contained more
degrees of freedom than conventional ones.} In this way the
superembedding encompasses properties of both the Green--Schwarz
and the Neveu--Schwarz--Ramond description of the superstring. In
the superembedding formulation the spinor coordinates
$\theta^{\underline\alpha}(\xi)$ of superstrings (and in general
of the superbranes) have worldvolume superpartners, auxiliary
commuting spinor variables $\lambda^{\underline\alpha}(\xi)$ which
have the properties of twistors. This is why initially such
formulations were called `twistor--like'. In a formulation of the
theory in terms of $n=2$ worldsheet superfields the conjugate
spinors $\lambda^{\underline\alpha}(\xi)$ and
$\bar\lambda^{\underline\alpha}(\xi)$ are complex and satisfy the
pure spinor condition\footnote{Pure spinors were introduced by
Cartan \cite{cartan} and more recently have been considered also
in \cite{budinich}.} $\lambda\Gamma^{\underline
m}\lambda=0=\bar\lambda\Gamma^{\underline m}\bar\lambda$ (which is
a part of the superembedding condition)
\cite{mario,berktwistor,tonintwistor}, and
$\theta^{\underline\alpha}(\xi)$ also has a fermionic
superpartner, an auxiliary field $\sigma^{\underline\alpha}$. Pure
spinors have been considered as auxiliary fields in
\cite{nilsson}. Later on, Howe derived superspace constraints in
$D=10$ SYM and $D=10,11$ supergravity theories as integrability conditions
along pure spinor lines \cite{Howelines}.

The problem of the covariant quantization of the superstring using
the `twistor--like' formulation, and in particular most suitable
$n=2$ supersymmetric pure spinor models, has been mainly assaulted
by Berkovits \cite{berkassault}.\footnote{Related, though rather
cumbersome, methods used Lorentz harmonics to insure covariance
\cite{nis,rakh,ig}.} Having started from superembedding techniques
in \cite{berktwistor}, he has recently arrived at a pure spinor
method of covariant quantization
\cite{berkpurespinor,berkrelating,purespectrum}.  The pure spinor
formalism has been also applied to superparticles, to a
supermembrane \cite{berkother} and to open superstrings
\cite{pershin}. In this long way a link with the original
superembedding formulation has been lost somewhere.

In this paper we find and restore this link. Since the
superembedding formulation is classically equivalent to the
Green--Schwarz formulation we therefore also find a relation
between the reparametrization invariant and $\kappa$--symmetric
Green--Schwarz superstring action and the action of Berkovits,
thus completing results earlier obtained in \cite{odatonin}. We do
this with the example of the heterotic string. Substantially the
Berkovits action is a complexified superembedding formulation
gauge--fixed in a conventional BRST manner.

A key point in establishing the relation with the Berkovits method
is to complexify the model. Note that this step does not double
the physical degrees of freedom since the complex conjugate fields
never appear in the action of the model. The complexification is
useful for two reasons, it allows one to treat $\lambda$ and
$\bar\lambda$ as independent fields and to perform the twisting of
an $n=(0,2)$ superconformal system \cite{twist} associated with
the superstring model. Namely, upon complexification, solving a
part of the superembedding condition and gauge fixing the
$\kappa$--symmetries involving $\bar\lambda$ and $\sigma$, one can
express them in terms of other fields of the model thus removing
$\bar\lambda$ and $\sigma$ as independent fields from the action.
In this stage the model is still invariant under
$\kappa$--symmetries acting on $\theta$ and $\lambda$ as well as
under the $n=(0,2)$ superdiffeomorphisms whose gauge fixing gives
rise to an associated $n=(0,2)$ superconformal system.

Twisting the $n=(0,2)$ superconformal system has two important consequences.
Firstly, it shifts the conformal weights of the fields with a non vanishing
$U(1)$ R---charge and makes the superconformal algebra of currents anomaly free
(i.e. its central charge vanishes). In particular, the central charge of the
system of superdiffeomorphism ghosts, which originally is $-6$, vanishes after
the twist. This shall allow us to drop from the action the superdiffeomorphism
ghost system which eventually would arise upon gauge fixing the
superdiffeomorphisms. Finally, the remaining $\kappa$--symmetries are gauge
fixed by a standard BRST recipe, i.e. by adding to the action the BRST
variation of a suitable `gauge fermion' and using as a BRST charge the twisted
superconformal charge with positive ghost number (positive R---charge before
the twist), as the twisting procedure prescribes. In this way one recovers the
Berkovits action with a simple BRST charge of the form $ Q_B = \oint
\lambda^{\underline \a} d_{\underline\a} $, where $d_{\underline\a}$ is a
fermionic constraint that acts as a covariant spinor derivative on target space
superfields. The cohomology of $Q_B$ has been proven to give the correct
superstring spectrum \cite{purespectrum}. Furthermore, the twisted
superconformal current with a negative ghost number reproduces a correct
composite `$b$'--field of the Berkovits approach. We thus show that the
Berkovits pure spinor formulation is a gauge fixed and twisted version of the
(complexified) superstring in the superembedding approach. This provides the
Berkovits method with a geometrical superdiffeomorphism invariant ground and
explains both the absence of the (super)diffeomorphism ghosts in the pure
spinor quantization prescription and the nature of its BRST charge and
antighost.

The paper is organized as follows. In Section~2 and ~3 we introduce the action
for the $D=10$ heterotic string in the superembedding formulation in terms of
$n=(0,2)$ worldsheet superfields \cite{berktwistor,tonintwistor}, analyze its
local symmetries (which include the $n=(0,2)$ superdiffeomorphisms and six
$\kappa$--symmetries) and demonstrate how it is related to the Green--Schwarz
action. Section~4 is devoted to a detailed analysis of the $n=2$ superconformal
structure of the action. In Section~5 we consider the procedure that leads from
the superembedding model to the pure spinor effective action by means of the
complexification, the gauge fixing of the local symmetries and the twisting of
the $n=2$ superconformal theory. In Section~6 we generalize the construction to
the case of a supergravity and super--Yang--Mills background. Finally, in
Section~7 we present conclusions and suggestions for further study.

\section{The $n=(0,2)$ superembedding formulation of the heterotic string}
\setcounter{equation}0

Let us start by assuming that the worldsheet of the heterotic
string be an $n=(0,2)$ supersurface parametrized by two bosonic
(light--cone) coordinates $\xi^{++},~\xi^{--}$ and two complex
conjugate fermionic coordinates $\eta^-,~\bar\eta^-$ collectively
defined as $z^M=(\xi^{++},\xi^{--},\eta^-,\bar\eta^-)$. Here each
$+$ and $-$ corresponds to the $d=2$ Lorentz group weight
$-{1\over 2}$ and $+{1\over 2}$ respectively. These become
{right}-- and {left}--sector conformal weights upon gauge fixing the
worldsheet superdiffeomorphisms (we shall call them `conformal
weights'). Note that `$--$' and `++' are the $d=2$ light--cone
vector indices and `$-,+$' are the Majorana spinor indices). The
worldsheet supersurface is embedded into an $N=1$, $D=10$ target
superspace parametrized by ten bosonic coordinates $x^{\underline
m}$ and by sixteen Majorana--Weyl fermionic coordinates
$\theta^{\underline\mu}$ collectively defined as $Z^{\underline
M}$, where $\underline{M}=
(\underline{m},\underline{\mu})$.\footnote{Underlined and non
underlined letters denote respectively target superspace and
worldsheet superspace indices. Letters from the beginning and from
the middle of the alphabet denote tangent superspace and
supermanifold indices respectively. Of course, in flat superspace,
curved and spinor indices may be identified. Also note that in our
notation the complex conjugate of the product of two fermions is
$(\psi_1\psi_2)^*=\bar\psi_1\bar\psi_2$.} The image of the
superworldsheet in the target superspace is described by the
$n=(0,2)$ superfields
\begin{equation}\label{image1}
X^{\underline m}(z)=x^{\underline m}(\xi)+\eta^-\chi^{\underline
m}_-(\xi)+\bar\eta^-\bar\chi^{\underline
m}_-(\xi)+i\eta^-\bar\eta^-\nu_{--}^{\underline m}(\xi)\,,
\end{equation}
\begin{equation}\label{image2}
\Theta^{\underline
\mu}(z)=\theta^{\underline\mu}(\xi)\,+\,\eta^-\lambda
^{\underline \mu}_-(\xi)\,+\,\bar\eta^-\bar\lambda^{\underline
\mu}_-(\xi)\,+\,i\eta^-\bar\eta^-\sigma_{--}^{\underline\mu}(\xi
)\,,
\end{equation}
whose leading components are the target superspace coordinates of
the heterotic string. For the model to have the correct number of
physical degrees of freedom, i.e. that of the heterotic string,
the higher components of (\ref{image1}) and (\ref{image2}) should
be auxiliary worldsheet fields. Therefore the superfields
(\ref{image1}) and (\ref{image2}) should be constrained in a
proper way. The constraint is the superembedding condition
\begin{equation}\label{sc}
E^{\underline a}_{-}(Z^{\underline M}( z))=0=\bar E^{\underline
a}_{-}(Z^{\underline M}( z))\,,
\end{equation}
where $E^{\underline a}_{-}$ and $\bar E^{\underline a}_{-}$ are
spinor components of the pullback onto the superworldsheet of the
vector component of the target space supervielbein one form
\begin{equation}\label{E}
E^{\underline A}=e^A(z) E_A^{\underline A}(Z(z))=
e^{++}E_{++}^{\underline A}\,\,+e^{--}E_{--}^{\underline
A}\,+\,e^{-}E^{\underline A}_-\,+\,\bar e^{-}\bar E^{\underline
A}_-\,,
\end{equation}
$e^A(z)$ being a worldsheet supervielbein one form and
$\underline{A}=(\underline{a},\underline{\a})$. For consistency
the supervielbeins should satisfy worldsheet and target space
supergravity constraints (see \cite{report} and references therein
for details).

In flat target superspace the superembedding condition on the
pullback of the superinvariant form \be\label{Pi} \Pi^{\underline
m}=dX^{\underline m}-\half d\Theta\G^{\underline
m}\Theta=e^{++}\Pi^{\underline m}_{++}+ e^{--}\Pi^{\underline
m}_{--}+e^{-}\Pi^{\underline m}_{-}+\bar e^{-}\bar\Pi^{\underline
m}_{-}\,, \ee where $d=e^A\,D_A$ is the worldsheet superspace
differential, reduces to
 \be\label{flat}
 \Pi^{\underline m}_{-}\equiv D_{-}X^{\underline
m}-\half D_{-}\Theta\G^{\underline m}\Theta=0,\quad
 \bar\Pi^{\underline m}_{-}\equiv \bar D_{-}X^{\underline
m}-\half\bar D_{-}\Theta\G^{\underline m}\Theta=0\,,
 \ee
whose integrability requires the twistor--like constraint
 \be\label{twist} \Pi^{\underline
m}_{--}=\half D_{-}\Theta\G^{\underline m}\bar D_-\Theta\,, \ee
and the pure spinor condition for the superfields $D_{-}\Theta$
and $\bar D_-\Theta$, that is
\be\label{PURE}
 D_{-}\Theta\G^{\underline m}D_-\Theta=0, \qquad \bar
D_{-}\Theta\G^{\underline
 m}\bar D_-\Theta=0\,.
\ee
 The worldsheet supercovariant derivatives act as follows
 on a superfield $\Phi^{(+p,-q)}(z)$ of conformal weights
$(+p,-q)$
 (being, by definition, the weights of its leading component)
$$ D_{-}\Phi^{(+p,-q)}
=\left({\partial\over{\partial\eta^-}}+\bar\eta^-
D_{--}\right)\Phi^{(+p,-q)}, ~ \bar D_{-}\Phi^{(+p,-q)}
=\left({\partial\over{\partial\bar\eta^-}}+\eta^-
D_{--}\right)\Phi^{(+p,-q)}\,,
$$
\bea\label{fermionic}
 D_{--}\Phi^{(+p,-q)}=\left(\partial_{--}\,+\,e^{++}_{--}\partial_{++}
 +p\;\partial_{++}e^{++}_{--}\right)\Phi^{(+p,-q)}\,, \quad
 D_{++}\Phi^{(+p,-q)} =\partial_{++}\Phi^{(+p,-q)}\,,
 \quad
 \eea
$e^{++}_{--}(\xi)$ being a Beltrami parameter (i.e. the only
nontrivial component of the worldsheet supervielbein (\ref{E})).

The superembedding condition (\ref{flat}) leads to the following
relations between the components of the superfields (\ref{image1})
and (\ref{image2})\footnote{Analogous relations also hold in the
curved supergravity background for the components of the target
space supervielbein pullbacks.}
\be\label{chi}
\chi^{\underline m}_--\half\theta\Gamma^{\underline m}\lambda_-=0,
\quad \bar\chi^{\underline m}_--\half\theta\Gamma^{\underline
m}\bar\lambda_-=0\,, \quad \nu^{\underline
m}_{--}-\half\sigma_{--}\Gamma^{\underline m}\theta=0\,,
 \ee
 \be\label{lgl}
\l_-\G^{\underline m}\l_-=0,\qquad \bl_-\G^{\underline m}\bl_-=0\,,
\ee \be \Pi^{\underline
m}_{--}|_{\eta,\bar\eta=0}-\half\l_-\G^{\underline m}\bl_-
=\partial_{--}x^{\underline m}-\half
\partial_{--}\theta\G^{\underline m}\theta-\half\l_-\G^{\underline m}\bl_-
=0,\label{twistor} \ee \be\label{tsgl}
(\partial_{--}\t+i\s_{--})\G^{\underline m}\l_-=0\,,\qquad
(\partial_{--}\t-i\s_{--})\G^{\underline m}\bl_-=0\,.
 \ee
Eqs. (\ref{chi})--(\ref{tsgl}), together with the local symmetries
of the model, will allow us to express the higher components of
the superfields (\ref{image1}) and (\ref{image2}) in terms of
$x^{\underline m}$ and $\theta^{\underline\alpha}$ and their
derivatives. Note that eq. (\ref{lgl}) is the pure spinor
condition for $\lambda^{\underline\alpha}_-$ and
$\bar\lambda^{\underline\alpha}_-$, while the twistor--like
condition (\ref{twistor}) implies one of the Virasoro constraints,
namely \be\label{virasoro--} \Pi^{\underline
m}_{--}\Pi_{--\underline m}=0\,. \ee The fact that the
supermebedding condition does not produce dynamical equations of
motion for $x^{\underline m}$ and $\theta^{\underline\alpha}$
allows one to construct the $n=(0,2)$ worldsheet superfield action
for the heterotic string \cite{berktwistor} in an $N=1$, $D=10$
supergravity background \bea \label{seaction}
I_{SE}&=&\int d^2\xi d\eta_-
d\bar\eta_-\left[P^{-}_{++\underline a}E^{\underline a}_{-}+{\bar
P}^{-}_{++\underline a}\bar E^{\underline a}_{-}
+\half{\bar\Psi}^{I}_+\Psi^{I}_+\right]+\half\int d^2\xi
\left[d\eta_- B_{++,-}
+d\bar\eta_- \bar B_{++,-}\right]\,.\nn\\
\eea
In this action the superfields
\bea\label{P}
P^{-}_{++\underline a}(z)&=&\rho^{-}_{++\underline
a}+\eta^-a_{++\underline a}+\bar\eta^-r_{++\underline
a}+i\eta^-\bar\eta^-\rho_{-++\underline a}\,,\nonumber\\
 {\bar P}^{-}_{++\underline a}(z)&=&\bar\rho^{-}_{++\underline
a}+\bar\eta^-\bar a_{++\underline a}+\eta^-\bar r_{++\underline
a}+i\eta^-\bar\eta^-\bar\rho_{-++\underline a}\,,
 \eea
are complex conjugate Lagrange multipliers whose variation
produces the superembedding condition (\ref{sc}) and, as a
consequence, the Virasoro constraint (\ref{virasoro--}). The
second Virasoro constraint is obtained by varying the action with
respect to the Beltrami parameter $e^{++}_{--}(\xi)$
(\ref{fermionic}). $\Psi^{I}_+(z)$ and $\bar\Psi^{I}_+(z)$
($I=1,\cdots,16)$ are heterotic fermion (anti)chiral superfields
(i.e. $\bar D_-\Psi^{I}_+=0= D_-\bar\Psi^{I}_+$), and $B_{++,-}$,
$\bar B_{++,-}$ are complex conjugate spin--vector components of
the pullback onto the superworldsheet of the NS--NS two--form
potential $B^{(2)}$. It can be shown that modulo the
superembedding condition and the $N=1$, $D=10$ supergravity
constraints, the superfields $B_{++,-}$ and $\bar B_{++,-}$ are
chiral and antichiral respectively. For instance, in flat target
superspace \be\label{Bflat} B^{(2)}=d X^{\underline m}\;
d\Theta\G_{\underline m}\Theta=\Pi^{\underline m}\;
d\Theta\G_{\underline m}\Theta\,, \ee \bea\label{Bflat-}
B_{++,-}&=&\Pi^{\underline m}_{++}\;D_{-}\Theta\G_{\underline
m}\Theta -\Pi^{\underline m}_{-}\;D_{++}\Theta\G_{\underline
m}\Theta\,,
\nonumber\\
\bar B_{++,-}&=&\Pi^{\underline m}_{++}\;\bar
D_{-}\Theta\G_{\underline m}\Theta -\bar\Pi^{\underline
m}_{-}\;D_{++}\Theta\G_{\underline m}\Theta\,, \eea and
\be
\label{db+} D_{-}B_{++,-}= D_{++}(\Pi^{\underline
m}_{-}\,D_-\Theta\Gamma_{\underline m}\Theta)- 2D_{-} \Pi^{\underline
m}_{-}\,D_{++}\Theta\Gamma_{\underline m}\Theta-2\Pi^{\underline
m}_{-}\,D_{++}\Theta\Gamma_{\underline m}D_-\Theta \,,
\ee
which vanishes when (\ref{flat}) is satisfied.

By construction the target superspace covariant action
(\ref{seaction}) is invariant under worldsheet
superdiffeomorphisms, even if it does not contain the worldsheet
supervielbeins (apart from the Beltrami parameter
$e^{++}_{--}(\xi)$), provided that the superbackground satisfies
the $N=1$, $D=10$ supergravity constraints. Furthermore, the Lagrange
multipliers $P^{-}_{++\underline a},{\bar P}^{-}_{++\underline a}$
must vary in a proper way to cancel the terms, proportional to the
superembedding condition, coming from the variation of $B_{++,-}$
and $\bar B_{++,-}$.

The superdiffeomorphism variations of the target superspace
coordinates $Z^{\underline M}(z)$ are \be\label{sdif} \d
Z^{\underline M} =D_- {\cal C}^{--} \bar D_-Z^{\underline
M}\,+\bar D_{-}{\cal C}^{--} D_{-}Z^{\underline M}\, +\,2 \,{\cal
C}^{--}D_{--}Z^{\underline M}+ 2c^{++} D_{++}Z^{\underline M}\,,
\ee where $c^{++}(\xi)$ is a parameter of the {left}--sector
bosonic reparametrizations and \be\label{C} {\cal
C}^{--}(z)=c^{--}(\xi)+\eta^-\bar\gamma^-(\xi)
+\bar\eta^-\gamma^-(\xi)+i\eta^-\bar\eta^- v(\xi)\,, \ee is an
unconstrained superfield parameter of the {right}--sector
superdiffeomorphisms. In what follows we shall denote the BRST
ghosts associated with the superdiffeomorphisms (\ref{sdif}) with
the same letters as in (\ref{C}), and denote the corresponding
antighosts by
 \be\label{Bg} {\cal
B}_{--}(z)=u_{--}(\xi)+\eta^-\beta_{---}(\xi)+\bar\eta^-\bar
\beta_{---}(\xi)+i\eta^-\bar\eta^-b_{----}(\xi)\,. \ee

The odd components of the superdiffeomorphisms (\ref{sdif}) replace
two independent $\kappa$--symmetry transformations of the Green--Schwarz
formulation. The remaining six non--manifest $\kappa$--symmetries
of the $D=10$ heterotic string action (\ref{seaction}) are
realized as follows \bea \label{hatk}
 \d Z^{\underline M}E_{\underline
M}^{\underline \alpha}&=&E_-\G^{\underline a}\bar E_-\,
(\G_{\underline a}K^{--})^{\underline \a}-2E_-^{\underline
\alpha}\,\bar E_{-}K^{--} -2\bar E_{-}^{\underline\alpha}\,E_-
K^{--}\,,~~ \d Z^{\underline M}E_{\underline M}^{\underline
a}=0\,, \eea where, among the sixteen fermionic superfield
parameters $K^{--}_{\underline \a}(z)$ only six are independent.
The action (\ref{seaction}) is also invariant under the following
local transformations of the Lagrange multipliers
\cite{berktwistor} \be \label{csym} \d P^{-}_{++\underline
a}=D_-\Xi^{---}_{++}\G_{\underline a} E_-\,,\quad \d{\bar
P}^{-}_{++\underline a}=\bar D_{-}\bar
\Xi^{---}_{++}\G_{\underline a}\bar E_-\,, \ee where among the
complex conjugate superfield parameters $\Xi^{---\underline
\a}_{++}(z)\;$ and $\;{\bar \Xi}^{---\underline \a}_{++}(z)\;$
only ten real are independent. The spinor supervielbein pullbacks
$E^{\underline\alpha}_-$ and $\bar E^{\underline\alpha}_-\;$
reduce to $D_-\Theta^{\underline\alpha}$ and $\bar
D_-\Theta^{\underline\alpha}$ in flat target superspace.

We now proceed with analyzing the action (\ref{seaction}) and the
superembedding condition in the flat superbackground and will
consider its coupling to $N=1$, $D=10$ supergravity and
super--Yang--Mills in Section 6.

\section{The Green--Schwarz action from superembedding}
\setcounter{equation}0

Let us first perform the integration over the fermionic
coordinates in the action (\ref{seaction}). Next we solve for the
auxiliary fields $\chi_-^{\underline m}$, $\bar\chi_-^{\underline
m}$ and $\nu_{--}^{\underline \alpha}$ by means of (\ref{chi}). Then,
eliminating the corresponding components of the Lagrange
multipliers (\ref{P}), using the equations of motion of
$\chi_-^{\underline m}$ and $\bar\chi_-^{\underline m}$ and
(\ref{csym}), we obtain \bea\label{compact} I_{SE}&=&\half\int
d^2\xi\left[\Pi^{\underline m}_{++}\Pi_{--\underline
m}+\half\Pi^{\underline m}_{++}\,D_{--}\,\t\G_{\underline m}\t
-\half\Pi^{\underline
m}_{--}\,D_{++}\,\t\G_{\underline m}\t \right.\nn\\
&+&\left.
i\bar\psi^{I}_+D_{--}\psi^{I}_++i\psi^{I}_+D_{--}\bar\psi^{I}_+\right]
\nonumber\\
&+&\int d^2\xi\left[ p_{++\underline m}(\Pi^{\underline
m}_{--}-\half\l_-\G^{\underline m}\bl_-)+r_{++\underline
m}\l_-\G^{\underline m}\l_- + {\bar r}_{++\underline
m}\bar\l_-\G^{\underline
m}\bar\l_-\right.\nonumber\\
 &+& \left.\rho^-_{++\underline
m}(\partial_{--}\t+i\s_{--})\G^{\underline m}\l_-+
 \bar\rho^-_{++\underline m}(\partial_{--}\t-i\s_{--})\G^{\underline
m}\bar\l_-\right]\,, \eea where \be \Pi^{\underline m}_{\pm\pm}\equiv
D_{\pm\pm}x^{\underline m}-\half D_{\pm\pm}\,\t\G_{\underline m}\t\,, \ee now
stands for the leading components of the superfields (\ref{Pi}),
$\psi^{I}=\Psi^{I}|_{\eta=\bar\eta=0}$,
$\bar\psi^{I}=\bar\Psi^{I}|_{\eta=\bar\eta=0}$, while $p_{++\underline
m}=2a_{++\underline m}+2\bar a_{++\underline m}-\Pi_{++\underline m}$.
Furthermore ${r}_{++\underline m}$, ${\bar r}_{++\underline m}$,
$\rho^-_{++\underline m}$ and $\bar\rho^-_{++\underline m}$ are the remaining
components of the Lagrange multipliers (\ref{P}).

If we now solve the equations of motion of $\l_-$, $\bar\l_-$ and
$\sigma_{--}$, and gauge fix the local symmetries remaining in
(\ref{csym}), we can set $r_{++\underline m}=\bar r_{++\underline
m}=\rho^-_{++\underline m}=\bar\rho^-_{++\underline m}=0$ and find
that \be\label{p} p_{++\underline m}=
e^{--}_{++}(\xi)\l_-\G_{\underline m}\bar\l_-\,. \ee Here
$e^{--}_{++}(\xi)$ is a worldsheet parameter which can be regarded
as the second component of the worldsheet vielbein, in addition to
the Beltrami parameter $e^{++}_{--}(\xi)$ in (\ref{fermionic}).

The superfield counterpart of eq. (\ref{p}) is \bea\label{calE}
D_-( P^{-}_{++\underline m}+\half\eta^-\Pi_{++\underline m
})-\Pi_{++\underline m}=\half{\cal
E}^{--}_{++}(z)D_-\Theta\Gamma_{\underline m}\bar D_-\Theta\,
~~{\rm and}~~c.c. \,, \eea where ${\cal E}^{--}_{++}(z)$ is a
superfield, containing $e^{--}_{++}(\xi)$ as the leading
component, which can be regarded as a super--Beltrami parameter
associated with the $n=(0,2)$, $d=2$ superdiffeomorphisms.

Substituting (\ref{p}) into (\ref{compact}), taking into account
that $(\l_-\G_{\underline
 m}\bar\l_-)^2=0$ because of the pure spinor condition, and using once
 again the twistor--like condition $\Pi_{--}^{\underline m}=\half
\l_-\G_{\underline
 m}\bar\l_-$ to eliminate $\lambda$ and $\bar\lambda$ from the action,
we get the Green--Schwarz action for the heterotic string in the
following form \be\label{gs} I_{GS}=\half\int
d^2\xi\left[e_{++}^me_{--}^n\Pi^{\underline m}_{m}\Pi_{n\underline
m}+\epsilon^{mn}\Pi^{\underline
m}_{m}\,\partial_{n}\,\t\G_{\underline
m}\t+e_{--}^m(\bar\psi^{I}_+\partial_m\psi^{I}_++\psi^{I}_+\partial_m
\bar\psi^{I}_+)\right]\,, \ee where \be\Pi^{\underline
m}_{m}=\partial_mx^{\underline
m}-\half\partial_{m}\,\t\G^{\underline m}\t\,,\ee and
$e_{++}^m=(1-e_{++}^{--}e_{--}^{++},\,e_{++}^{--})$,
$e_{--}^m=(e_{--}^{++}, 1)$ (notice that $\det\,e^m_a=1$).

Let us note that the first integral in the action (\ref{compact})
coincides with the Green--Schwarz action (\ref{gs}) when in the
latter the {right}--sector diffeomorphisms
$\d\xi^{--}=c^{--}(\xi)$ are gauge fixed by imposing (locally) the
conformal gauge $e^{--}_{++}(\xi)=0$. Therefore, the
superembedding action is basically the Green--Schwarz action gauge
fixed in the {right}--sector plus the terms which ensure the
superembedding condition and the full {left}--right
(super)diffeomorphism invariance of the construction. We have thus
demonstrated the classical equivalence of the superembedding and
the Green--Schwarz formulation of the heterotic superstring.

\section{The $n=(0,2)$ superconformal structure of the superembedding
action} \setcounter{equation}0 Let us now analyze the form of the
first--class constraints generating the $n=(0,2)$, $d=2$
superdiffeomorphisms which comprise the $U(1)$ current $j(\xi)$ of
R--symmetry, two local supersymmetry currents $G(\xi)$ and $\bar
G(\xi)$, and the energy momentum tensor $T(\xi)$ associated with
the {right}--sector bosonic diffeomorphisms.\footnote{To simplify
notation we have omitted to specify the conformal weights of the
currents $j, G,\bar G, T$, which are 1, 3/2, 3/2, and 2,
respectively.} These are conserved currents of matter fields.
Furthermore, they are first class constraints of the model which
form an $n=(0,2)$, $d=2$ supermultiplet that can be cast into the
superfield \bea\label{Jg} J_{--}&=&j +i\eta^- G -i\bar\eta^- \bar
G - {2i}\eta^-\bar\eta^- T \,, \eea whose explicit form can be
found by the standard Noether procedure. To this end one considers
the variation of the action (\ref{seaction}) under $U(1)$
R--symmetry transformations of the Grassmann--odd coordinates
$\delta\eta=i\phi\eta,~\delta\bar\eta=-i\phi\bar\eta$ and regards
$D_{++}$ as the time derivative in order to identify the
corresponding conjugate momenta of $X^{\underline m}$ and
$\Theta^{\underline\alpha}$. Note that only the `Wess--Zumino'
$B^{(2)}$--term (\ref{Bflat-}) of the action contributes to the
definition of the supercurrent which, up to a square of the
superembedding condition, has the form
 \bea\label{J} J_{--}&=&i(D_-\Theta W_--\bar D_{-}\Theta\bar
W_-)+iD_-X^{\underline m}\bar D_-X_{\underline m}- {i\over 4}
D_-\Theta \G_{\underline
m}\Theta\bar D_-\Theta\G_{\underline m}\Theta \nonumber\\
 &=&i(D_-\Theta W_--\bar D_{-}\Theta \bar W_-)+
{i\over 4}\Pi^{\underline m}_-\,\bar D_-\Theta\G_{\underline
m}\Theta -{i\over 4}\bar\Pi^{\underline m}_{-}\,D_-\Theta
\G_{\underline m}\Theta\,. \eea Here $W_{-\underline\alpha}\approx
0$ and $\bar W_{-\underline\alpha}\approx 0$ are complex conjugate
chiral superfield constraints\footnote{The symbol $\approx$
indicates that the Hamiltonian constraints are in general
satisfied in a weak sense.} containing the canonical momenta of
the components of $\Theta^{\underline\alpha}$ and
$\bar\Theta^{\underline\alpha}$. Namely \bea\label{W}
W_{-\underline\alpha}&=&\omega_{-\underline\alpha}+
\eta^-\left[d_{--\underline\a}-iD_{--}\tau_{\underline\a}
+\half(\Pi_{--}^{\underline m}-\half\lambda_-\Gamma_{\underline
m}\bar\lambda_-)(\Gamma^{\underline
m}\theta)_{\underline\a}\right]
+\eta^-\etab^-D_{--}\omega_{-\underline\alpha}\,,\nn\\
\bar W_{-\underline\alpha}&=&\bar\omega_{-\underline\alpha}+
\bar\eta^-\left[d_{--\underline\a}+iD_{--}\tau_{\underline\a}
+\half(\Pi_{--}^{\underline m}-\half\lambda_-\Gamma_{\underline
m}\bar\lambda_-)(\Gamma^{\underline
m}\theta)_{\underline\a}\right]
-\eta^-\etab^-D_{--}\bar\omega_{-\underline\alpha}\,,\nn\\
\eea where $\omega_{-\underline\alpha}$ and
$\bar\omega_{-\underline\alpha}$ are the conjugate momenta to
$\lambda_{-\underline\alpha}$ and
$\bar\lambda_{-\underline\alpha}$ respectively and
$\tau_{\underline\alpha}$ is the conjugate momentum to the
auxiliary field $\sigma_{--}^{\underline\alpha}$ of
(\ref{image2}). The field \be
d_{--\underline\a}=p_{--\underline\a}-\half\Pi^{\underline
m}_{--}(\G_{\underline
m}\t)_{\underline\a}-{1\over8}(D_{--}\t\G^{\underline
m}\t)(\G_{\underline m}\t)_{\underline\a}\approx 0\,, \ee which
satisfies the Poisson brackets \be
\{d_\alpha,d_\beta\}=-\Gamma^{\underline
m}_{\underline\alpha\underline\beta}\Pi_{--\underline m}\,, \ee is
the usual Green--Schwarz supercovariant momentum.

The component fields of $J_{--}$ are given by
\bea\label{components} j
&=&J_{--}|_{\eta=\etab=0}=i(\lambda^{\underline\a}_-\,
\omega_{-\underline\a}
-\bar\lambda^{\underline\a}_-\,\bar\omega_{-\underline\a})+\ldots\,,\\
G
&=&-iD_-J_{--}|_{\eta=\etab=0}=\lambda^{\underline\a}_-\,(d_
{--\underline\a} -i\,D_{--}\tau_{\underline\a}) +
(i\sigma^{\underline\a}_{--}-
D_{--}\theta^{\underline\a})\,\bar\omega_{-\underline\a}+\ldots\,, \\
\bar G &=&i\bar
D_{-}J_{--}|_{\eta=\etab=0}=\bar\lambda^{\underline\a}_{-}\,
(d_{--\underline\a}
 +i\, D_{--}\tau_{\underline\a})
- (i\sigma^{\underline\a}_{--}+
D_{--}\theta^{\underline\a})\,\omega_{-\underline\a}+\ldots,\\
T &=&{i\over 4}( D_{-}\bar D_--\bar D_-
D_{-})J_{--}|_{\eta=\etab=0}\nn\\
&=&-\omega_{-\underline\a}D_{--}\lambda^{\underline\a}_- +\half
D_{--}(\lambda^{\underline\a}_-
\,\omega_{-\underline\a})-\bar\omega_{-\underline\a}D_{--}\bar\lambda^
{\underline\a}_- +\half D_{--}(\bar\lambda^{\underline\a}_-\,
\bar\omega_{-\underline\a})\nn\\
&&+D_{--}\theta^{\underline\a}\,d_{--\underline\a}
+\sigma^{\underline\a}_{--}\,D_{--}\tau_{\underline\a} -
\half\Pi^{\underline m}_{--}\,\Pi_{--\underline
m}+\ldots\,,\label{T}
\eea
where `$\ldots$' denotes terms proportional to the superembedding
constraints (\ref{chi})--(\ref{tsgl}).

In the superconformal gauge ${\cal E}^{--}_{++}(z)=0$,
$e_{--}^{++}(\xi)=0$, the currents (\ref{components})--(\ref{T})
generate the {right}--sector $n=(0,2)$ super Virasoro algebra
realized on the matter fields. Thus, in the superconformal gauge,
the $n=(0,2)$ superembedding formulation of the heterotic
superstring is an interacting $n=(0,2)$, $d=2$ superconformal
model. Note that this model is still invariant under the six
$\kappa$--symmetries (\ref{hatk}).

In the next section we shall use these symmetries to gauge fix the
superembedding action in such a way that it would transform into
the Berkovits action.

\section{Complexified superembedding and Berkovits action}
\setcounter{equation}0 To quantize the action (\ref{compact}) one
should gauge fix the worldsheet superdiffeomorphisms and the local
$\kappa$--symmetries (\ref{hatk}). The gauge fixing of the local
$\kappa$--symmetries in a way which preserves a $U(4)$ subgroup of
the Lorentz group $SO(1,9)$, resulting in a consistent
semi--covariant quantization of the superstring, has been
performed in \cite{berktwistor}. However, our present goal is to
recover from the superembedding the Berkovits pure spinor
formulation, which provides one with a $D=10$ super--Poincar\'e
covariant quantization. Therefore, we should proceed in a
different way.

The Berkovits action (see eq. (\ref{B})) does not contain the pure
spinor $\bl$. Furthermore, under BRST transformations the `real'
spinor $\theta$ acquires a complex variation proportional to
$\lambda$, therefore this action is strictly speaking complex. So
we shall also `complexify' the superembedding model, i.e. consider
the superfields (\ref{image1}) and (\ref{image2}) as complex ones.
This procedure makes sense at the quantum level where the fields
are operators and the Fock space has in general an indefinite
metric.

The complexification has two useful consequences. On the one hand,
it allows us to treat $\l$ and $\bl$ in an asymmetric fashion and,
in particular, to use available local symmetries (\ref{hatk}) for
eliminating the pure spinor $\bl$ from the action (\ref{compact}).
On the other hand, it allows us to twist the $n=(0,2)$
superconformal model which arises upon gauge fixing the worldsheet
superdiffeomorphisms.

Let us begin with expressing $\bar\lambda^{\underline\alpha}_-$ in
terms of other fields of the model. To this end let us consider
the $U(5)$ subgroup of the (Wick rotated) Lorentz group $SO(10)$,
under which an $SO(10)$ vector $V^{\underline a}
\equiv(V^r,V_r),\,r=1,\ldots,5$, decomposes as a $\bf 5$--irrep
$V^r$ and a $\bf\bar 5$--irrep $V_r=V^{r\ast}$ of $U(5)$. An
$SO(10)$ (chiral) spinor
$\phi^{\underline\a}\equiv(\phi^0,\phi_{[rs]},\phi^r)$ decomposes
as $(\bf{1},\overline{\bf 10},\bf{5})$ of $U(5)$. Let
$v^0_{\underline\a}$ be a constant $1\times16$ $c$--number matrix
that extracts from $\phi^{\underline\a}$ the $U(5)$ singlet
$\phi^0=v^0_{\underline\a}\phi^{\underline\a}$. Notice that
$v^0_{\underline\a}$ satisfies the pure spinor condition
$(v^0\Gamma^{\underline a}v^0)=0$.\footnote{It would be of
interest to consider the possibility of promoting the constant
$v^0_{\underline\a}$ to a component of a worldsheet harmonic
matrix parametrizing a coset space ${SO(10)\over U(5)}$, thus
making $v^0_{\underline\a}$ an auxiliary harmonic field (for the
use of harmonics in the superembedding description of superstrings
see e.g. \cite{igor}).} Then one may define the pure spinor
\be\label{Y}
Y^-_{\underline\a}={v^0_{\underline\a}\over{v^0D_-\Theta}}\,,
\qquad Y^-D_-\Theta\equiv 1\,, \ee and construct the projector
\be\label{calP} {\cal
P}^{\underline\a}{}_{\underline\b}=\half(\G^{\underline m}
Y)^{\underline\a}(D_-\Theta\G_{\underline m})_{\underline\b}\,,
\quad {\cal P}{\cal P}={\cal P}\,, \quad (1-{\cal P})(1-{\cal
P})=(1-{\cal P})\,. \ee Since $D_-\Theta$ is a pure spinor, we
have \be\label{calPid} {\cal P}D_-\Theta= 0\,, \qquad
D_-\Theta\Gamma^{\underline m}(1-{\cal P})=0\,. \ee Note that
$\Tr\, {\cal P}=5$ and $\Tr\, (1-{\cal P})=11$, hence ${\cal P}$
and $1-{\cal P}$ project the $16$--dimensional spinor space onto
$5$-- and $11$--dimensional subspaces respectively. In particular,
(\ref{calPid}) reflects the fact that the pure spinor $D_-\Theta$
has 11 independent components.

As $Y^-$ and ${\cal P}$ contain $v^0_{\underline\a}$, they are
non--covariant, nevertheless they appear only at an intermediate
step as a technical tool in the passage from the covariant
superembedding to the covariant Berkovits action. In other words,
though we shall fix part of the $\kappa$--symmmetries (\ref{hatk})
in a way which breaks $SO(10)$ down to $U(5)$, the complete gauge
fixing of all local symmetries will restore the Lorentz
covariance.

Consider now the pure spinor condition for $\bar D_-\Theta$ whose
first component is $\bar\lambda_-$ \be \label{hatblgbl} \bar
D_{-}\Theta\G^{\underline
 m}\bar D_-\Theta=0\,.
\ee As it has been discussed, this condition implies that only
eleven of the sixteen components of $\bar D_-\Theta$ are
independent. Since any superfield $S^{\underline\alpha}$ such that
$S={\cal P}S$ identically satisfies the pure spinor condition
$S\G^{\underline m}S=0$, one can immediately see that the five
components of $\bar D_-\Theta$ `eliminated' by (\ref{hatblgbl})
belong to the projected part $(1-{\cal P}){\bar D_-\Theta}$. Also,
the six $\kappa$--transformations (\ref{hatk}) acting on $\bar
D_-\Theta$ affect only its projected part $(1-{\cal P})\bar
D_-\Theta$, because ${\cal P}\d\Theta\equiv 0$. Since, as we
mentioned, $(1-{\cal P}){\bar D_-\Theta}$ has eleven independent
components, five of which are zero due to (\ref{hatblgbl}) and the
remaining six transform under $\kappa$--symmetry, we can impose
the condition \be\label{Kgauge} (1-{\cal P}){\bar D_-\Theta}=0\,,
\ee which gauge fixes the part of $\kappa$--symmetry (\ref{hatk})
acting on ${\bar D_-\Theta}$. By
 (\ref{twist}) this condition is solved as
\be\label{sol}
{\bar D_-\Theta}^{\underline\a}=\Pi^{\underline m}_{--}
(\G_{\underline m}Y^-)^{\underline\a}\,,
\ee
which for the components of ${\bar D_-\Theta}$ implies
\be \label{95} \bar
\lambda^{\underline\alpha}_-=\Pi^{\underline m}_{--}
(\G_{\underline m}Y^-)^{\underline\a}|_{\eta=\bar\eta=0}\,, \ee
 and (due to (\ref{calPid}))
\be \label{96} (D_{--}\t-i\s_{--})^{\underline\a}=2{\cal
P}^{\underline\a}{}_{\underline\b}D_{--}\theta^{\underline\b}
+D_{--}(\chi^{\underline m}_--\theta\Gamma^{\underline
m}\lambda_-)(\G_{\underline m} Y^-)^{\underline\a})\,, \ee or
taking into account (\ref{chi}) \be\label{961}
\s_{--}^{\underline\a} =-i[(1-2{\cal
P})D_{--}\t]^{\underline\a}\,. \ee In (\ref{96}) and (\ref{961})
${\cal P}$ and $Y^-$ stand for the leading components of the
superfields (\ref{calP}) and (\ref{Y}), that below we shall denote
with the same symbol. We thus expressed
$\bar\lambda^{\underline\alpha}_-$ and
$\sigma^{\underline\alpha}_{--}$ in terms of
$\lambda^{\underline\alpha}_-$ and (the derivatives of)
$\theta^{\underline\alpha}$ and $x^{\underline m}$.

Note that the gauge fixing condition (\ref{95}) and the constraint
 $\bar\omega_{-\underline\alpha}\approx 0$ on the momentum
$\bar\omega_{-\underline\alpha}$, conjugate to
$\bar\lambda_{-}^{\underline\alpha}$, introduced in (\ref{W}),
form a canonical pair of second class constraints under the
Poisson brackets, i.e. $[\bar
\lambda^{\underline\alpha}_--\Pi^{\underline m}_{--}
(\G_{\underline
m}Y^-)^{\underline\a},\bar\omega_{-\underline\beta}]
=\delta^{\underline\alpha}_{\underline\beta}$. It follows that
$\bar\omega_{-\underline\alpha}$ can be considered to vanish in
the strong sense and therefore can be dropped in the definition of
the $n=(0,2)$ supercurrent components
(\ref{components})--(\ref{T}). The same reasoning applies to
$\sigma^{\underline\alpha}_{--}$ and its conjugate momentum
$\tau_{{\underline\alpha}}$ which can be strongly put to zero.

The expressions (\ref{95}) and (\ref{961}) identically satisfy the
superembedding conditions (\ref{tsgl}). Then, substituting
(\ref{95}) into the twistor--like constraint (\ref{twistor}), we
reduce it to \be \label{97} \half\,Y^-\G^{\underline
m}\G_{\underline n}\l_-\,\Pi_{--}^{\underline n}=0\,. \ee Apart
{}from the pure spinor condition $\lambda_-\G^{\underline
m}\lambda_-=0$, which we further assume to be satisfied in the
strong sense,\footnote{In a variation of the pure spinor
quantization method considered in \cite{senza} the pure spinor
constraint has been handled as a relaxed condition.} this is the
only constraint which remains in the model. So, upon substituting
the expression for $\bar\lambda_-$ (\ref{95}) into the action
(\ref{compact}), we have \bea\label{compacti}
 I_{SE}&=&\half\int
d^2\xi\left[\Pi^{\underline m}_{++}\Pi_{--\underline
m}+\half\Pi^{\underline m}_{++}\,D_{--}\,\t\G_{\underline m}\t
-\half\Pi^{\underline
m}_{--}\,D_{++}\,\t\G_{\underline m}\t \right.\nn\\
&+&\left.
\bar\psi^{I}_+D_{--}\psi^{I}_++\psi^{I}_+D_{--}\bar\psi^{I}_+\right]
+\half\int d^2\xi\, p_{++\underline m}\,(Y\G^{\underline
m}\G_{\underline n}\l\,\Pi_{--}^{\underline n})\,,
\eea
or a la Siegel \cite{alaSiegel}
\bea\label{compactis} I_{SE}&=&\int
d^2\xi\left[\half\partial_{++}x^{\underline m}D_{--}x_{\underline
m}+p_{--}\partial_{++}\t - d_{--}\partial_{++}\t+
\bar\psi^{I}_+D_{--}\psi^{I}_++\psi^{I}_+D_{--}\bar\psi^{I}_+\right]
\nonumber\\
&+&\half \int d^2\xi \, p_{++\underline m}\,(Y\G^{\underline
m}\G_{\underline n}\l\,\Pi_{--}^{\underline n})\,.
\eea The action (\ref{compacti}), and therefore also
(\ref{compactis}), is invariant under the $n=(0,2)$
superdiffeomorphisms and under the following $\kappa$--symmetry
transformations (\ref{hatk}) remained upon gauge fixing
$\bar\lambda_-$ and $\sigma_{--}$ \bea\label{residual}
\delta\theta^{\underline\alpha}
&=&(\delta_{\underline\beta}^{\underline\alpha}-
\lambda^{\underline\alpha}Y_{\underline\beta})\Gamma_{\underline
n}^{\underline{\beta\gamma}}\tilde\kappa^{--}_{\underline\gamma}
(\lambda\Gamma^{\underline n}\Gamma_{\underline m}Y)\Pi^{\underline m}_{--}\,, \\
\delta\lambda_-^{\underline\alpha}
&=&(\delta_{\underline\beta}^{\underline\alpha}-
\lambda^{\underline\alpha}Y_{\underline\beta})\Gamma_{\underline
n}^{\underline{\beta\gamma}}(\lambda\Gamma^{\underline
n}\Gamma_{\underline m}Y) [\Pi^{\underline
m}_{--}\tilde\mu^-_{\underline\gamma}
-(\lambda_-\Gamma^{\underline
m}D_{--}\theta)\tilde\kappa^{--}_{\underline\gamma}]\,, \\
\delta p_{++\underline m}&=&-\d\Pi_{++\underline m}=
\partial_{++}\theta\Gamma_{\underline m}\delta\theta\,, \eea
where
$\tilde\kappa^{--}_{\underline\gamma}=\kappa^{--}_{\underline\gamma}-
Y_{\underline\gamma}(\lambda\kappa^{--})$ and
$\tilde\mu^-_{\underline\gamma}=(\delta_{\underline\gamma}^{\underline
\alpha}-
Y_{\underline\gamma}\lambda^{\underline\alpha})D_-
K^{--}_{\underline\alpha}|_{\eta=\bar\eta=0}$ are components of
the superfield $\kappa$--symmetry parameter (\ref{hatk}).

Let us now compare (\ref{compactis}) with the Berkovits
action \be\label{B} I_B=\int
d^2\xi\left[\half\partial_{++}x^{\underline
m}\partial_{--}x_{\underline
m}+p_{--}\partial_{++}\theta+\omega_{--}\partial_{++}\lambda+
\bar\psi^{I}_+\partial_{--}\psi^{I}_++\psi^{I}_+\partial_{--}\bar
\psi^{I}_+
 \right]\,.
\ee
In this action, which describes a free conformal
system, $\lambda$ and $\omega_{--}$ have conformal weights 0 and
1, and ghost numbers 1 and $-1$, respectively, while in
(\ref{compactis}) $\lambda_-$, as well as its conjugate momentum
$\omega_-$, has conformal weight $\half$ and ghost number 0.

Therefore, as a first step in establishing the relationship between the actions
(\ref{compactis}) and (\ref{B}), we should change the conformal weights of
$\lambda_-$ and $\omega_-$ and endow them with appropriate ghost numbers. The
complexification allows one to do this via `twisting'. Indeed, before the
complexification the complex conjugate Grassmann coordinates $\eta^-$ and
$\bar\eta^-$ must have the same weight fixed to be $-\half$ because of the
relation $D_-\bar D_-+\bar D_-D_-=2D_{--}$. However, after the complexification
$\eta$ and $\bar\eta$ become independent variables, and one can choose their
weights to be respectively $-\half+w_0$ and $-\half-w_0$, with arbitrary $w_0$.
Choosing $w_0=\half$ one sees, by eqs. (\ref{image2}) and (\ref{W}), that
$\lambda$ acquires weight 0 and $\omega$ weight 1. Similarly, the supersymmetry
ghosts $\gamma_-$ and $\bar\gamma_-$ (\ref{C}) change their weights from
$-\half$ and $-\half$ to 0 and $-1$, respectively. As far as the change of the
ghost number is concerned, to turn the `matter' fields $\omega_-$ and
$\lambda_-$ into the ghost system with conformal weights (1,0) and ghost
numbers $(-1,1)$, one can use the ghost field $\gamma^-$ and make the field
redefinition $\lambda=\gamma^-\lambda_-$ and $\omega_{--}= {1\over
\gamma^-}\omega_-$. Actually, this can be done with the same
result before or after the twist.

In more precise terms the twisting procedure consists in the
following. One should first gauge fix the $n=(0,2)$
superdiffeomorphisms (\ref{sdif}) of the action (\ref{compactis})
to reduce it to an $n=(0,2)$ superconformal action. To this end,
we impose as gauge fixing that both the Beltrami parameter
corresponding to the {left}--sector diffeomorphisms and the one
corresponding to the {right}--sector superdiffeomorphisms vanish,
i.e. \bea e^{++}_{--}=0\,,&&{\cal E}^{--}_{++}=0\,. \eea This gauge
fixing requires the introduction of the system of
superdiffeomorphism ghosts (\ref{C}) and (\ref{Bg}).

In what follows we shall not consider the {left} sector
diffeomorphism ghosts, whose treatment is assumed to follow the
standard BRST procedure for the bosonic string, and will
concentrate on the {right}--moving (supersymmetric) sector of the
model whose quantization is problematic.

The system of the {right}--sector $n=(0,2)$ superdiffeomorphism
ghosts (\ref{C}) and (\ref{Bg}) consists of the
 fermionic ghost pairs $(b,~c^{--})$ of weights $(2,-1)$ and $(u,~v)$
 of weights (1,0), associated,
 respectively, with the {right}--sector bosonic diffeomorphisms and the $U(1)$
R--symmetry,
 and of bosonic ghost pairs $(\beta,~\gamma^-)$ and
 $(\bar\beta,~\bar\gamma^-)$ of weights $({3\over 2},-\half)$
 associated with the local $n=(0,2)$ supersymmetries.
 The $n=(0,2)$ ghost currents
\bea
j_{gh}&=&i(\beta\gamma - \bar\beta\bar\gamma)\,,\\
G_{gh}&=&-i\beta(v+i\partial c)-i\bar\gamma(b+i\partial u)\,,\\
\bar G_{gh}&=&i(b-i\partial u)\gamma +i\bar\beta(v-i\partial c)\,,\\
T_{gh}&=&2\left[i (\partial\beta)\gamma
+i\bar\beta\partial\bar\gamma +(\partial c)b +(\partial
u)v\right]\,, \eea should be added to the matter currents
(\ref{components})--(\ref{T}) which, upon the elimination of
$e^{++}_{--}$, $\bar\lambda_-$, $\sigma_{--}$ and their momenta,
take the form \bea\label{components1}
j &=&i\lambda^{\underline\a}_-\omega_{-\underline\a}\,,\\
G &=&\lambda^{\underline\a}_-\,d_{--\underline\a}\,, \\
\bar G &=&\Pi^{\underline m}_{--}\, Y^-\G_{\underline m}\, d_{--}
- 2\omega_{-}(1-{\cal P})\partial_{--}\theta\,,\\
T &=&-\omega_{-\underline\a}\partial_{--}\lambda^{\underline\a}_-
+\half
\partial_{--}(\lambda^{\underline\a}_-
\,\omega_{-\underline\a})+\partial_{--}\theta^{\underline\a}\,
d_{--\underline\a} -\half\Pi^{\underline
m}_{--}\,\Pi_{--\underline m}\,.\label{T1} \eea The currents
$j+j_{gh}$, $G+G_{gh}$, $\bar G+\bar G_{gh}$ and $T+T_{gh}$ form
the $n=(0,2)$ superconformal algebra of the matter + ghost system.

The twist of the $n=2$ superconformal algebra \cite{twist}
consists in shifting the stress--energy tensor $T$ as follows
\bea
T&\ar&T\,'=T+{i\over 2}\p j\,.
\eea
This corresponds to adding a
charge at infinity and has two important consequences:
\begin{itemize}
\item[i)] the conformal weight $w$ of any field $\phi$ with $R$--charge
$q$ gets
shifted \bea \label{twistcharge} w&\ar&w\,'=w-\half q\,,
\eea
\item[ii)] whatever the central charge was, it
vanishes after the twist.
\end{itemize}

For instance, after the twist the conformal weights of the pairs
$(\beta,~\gamma^-)$ and $(\bar\beta,~\bar\gamma^-)$ become (1,0)
and $(2,-1)$ respectively, and the central charge of the ghost
part of the $n=(0,2)$ superconformal generators, which was $-6$,
vanishes after the twist. Because of this, and since the ghost
sector is completely decoupled from the matter fields, we can
exclude the superdiffeomorphim ghosts from the consideration and
neglect them hereafter. This is the reason why the
superdiffeomorphism ghosts and, in particular, the $(b,c)$ system,
are not present in the pure spinor formalism.

We are thus left with the `matter' part of the $n=(0,2)$
superconformal system, in which the conformal weights get shifted
{}from $(\half,~ \half)$ to $(1,0)$, i.e.
$(\omega_-,~\lambda_-)~\rightarrow ~ (\omega_{--},~\lambda)$.

Then, according to \cite{twist}, after the twist one can
reinterpret $i\,j $ as the ghost number current with conformal
weight one and zero ghost number. Thus $\lambda$ acquires the
ghost number one and $\omega_{--}$ becomes a field with
$n_{gh}=-1$.\footnote{This is essentially the same as to say that
one has made the field redefinition of $\lambda$ and $\omega$ with
the use of the ghost $\gamma$.} As a consequence, $G =\lambda
d_{--}$ becomes a conserved current with $w=1$ and $n_{gh}=1$ and
can be identified with a BRST current whose associated BRST charge
\be\label{QB} Q_B=\oint \lambda d_{--}\,, \ee is nilpotent, since
$GG =0$ in virtue of the pure spinor property of $\lambda$. $\bar
G$ acquires conformal weight $2$ and ghost number $-1$ and is
therefore interpreted as the composite antighost field
\be\label{b} b_B=\half \bar G=\half \Pi^{\underline m}_{--}
\,Y\G_{\underline m}\, d_{--} - \omega_{--}(1-{\cal
P})\partial_{--}\theta\,. \ee We have thus shown that the twisting
of the $n=(0,2)$ superconformal algebra reproduces the BRST
current and the antighost field $b_B$ of the Berkovits pure spinor
formulation \cite{berkrelating}. In particular, the antighost
$b_B$, which is required for the construction of higher genus
amplitudes, satisfies the Poisson bracket anticommutation relation
\be\label{Qb} \{Q_B,b_B(\xi)\}=2T'(\xi)\,. \ee Though $b_B$ is not
Lorentz invariant, a consequence of the non--invariance of $Y$ and
${\cal P}$, its variation under the constant Lorentz
transformations $\Lambda_{\underline{mn}}$ is BRST exact
\be\label{Lb} \delta b_B=\half \{Q_B,\Pi_{--}^{\underline
m}\omega_{--}(1-{\cal P})\Gamma_{\underline m}\,\delta
Y\}\,+{1\over 16}
\Lambda_{\underline{mn}}\{Q_B,(Y\Gamma^{\underline m}
d_{--})\,(Y\Gamma^{\underline n} d_{--})\}\,, \ee where $\delta
Y_{\underline\alpha}={1\over 4}\Lambda_{\underline{mn}} (Y
\Gamma^{\underline{mn}})^{\underline\beta}
\;(\delta^{\underline\beta}_{\underline\alpha}-
\lambda^{\underline\beta}Y_{\underline\alpha})$.
 Eqs. (\ref{Qb}) and (\ref{Lb}) point to the fact that the physical
amplitudes should remain Lorentz covariant despite the
non--invariance of $b_B$.

In view of the above reasoning we identify $Q_B$ with the BRST
charge of the quantized theory and use it to gauge fix the
remaining $\kappa$--symmetries (\ref{residual}) of the action
(\ref{compactis}). To this end we introduce the gauge fermion
\be\label{fermi} {\cal F}={\cal F}_1+{\cal F}_2=\half
p_{++\underline m}\,d_{--}\Gamma^{\underline
m}Y+\omega_{--}(1-{\cal P})\partial_{++}\theta\,, \ee and add to
the action (\ref{compactis}) the gauge fixing term
\bea\label{fixing}
I_{gf}&=&\int d^2\xi\;\{Q_B,{\cal F}\}\nn\\
&=& \int d^2\xi\left[-\half p_{++\underline m}\,(Y\G^{\underline
m}\G_{\underline n}\l\,\Pi_{--}^{\underline n})+d_{--}{\cal
P}\partial_{++}\theta + d_{--}(1-{\cal P})\partial_{++}\theta
+\omega_{--}\partial_{++}\lambda\right]\,, \nn\\
\eea where we have used the BRST variations
$$\{Q_B,p_{++\underline
m}\}=-\partial_{++}\theta\Gamma_{\underline m}\lambda\,,\quad
\{Q_B,d_{--}\}=-\lambda\Gamma_{\underline n}\Pi^{\underline
n}_{--}\,,\quad
$$
$$
\{Q_B,\omega_{--}\}=d_{--}\,,\quad \{Q_B,\theta\}=\lambda\,.
$$
The resulting action $I_{SE}+I_{gf}$ is just the Lorentz covariant
Berkovits action (\ref{B}). The first term in (\ref{fixing})
cancels the residual superembedding term of (\ref{compactis}), the
second and third terms cancel $d_{--}\partial_{++}\theta$ and the
last one reproduces the kinetic contribution to the pure spinor
ghost $\lambda$.

Note that in \cite{odatonin} the second term of (\ref{fixing}) has
been added to the Green--Schwarz action `by hand' to render it
invariant under the BRST transformation generated by $Q_{B}$, and
the gauge fermion introduced therein was the ${\cal F}_2$ part of
(\ref{fermi}). In the superembedding formulation all the required
terms naturally appear in the BRST--exact gauge fixing part of the
superstring action.

We have thus shown that the Berkovits action is the gauge fixed
action of the complexified and twisted $n=(0,2)$ superembedding
formulation of the superstring. Therefore, since the latter is
classically equivalent to the Green--Schwarz formulation, we have
established the relationship between the Green--Schwarz and the
pure spinor superstring action.

An interesting problem, which should yet be understood, is the
relation between the Berkovits BRST charge (\ref{QB}), which
appears at the stage of twisting, and a conventional BRST charge
which one should construct when trying to quantize the
superembedding action (\ref{compacti}) directly. The conventional
BRST charge of the model should include all the constraints
generating the $n=(0,2)$ superconformal algebra
(\ref{components1}) and the $\kappa$--symmetry transformations
(\ref{residual}) and has the form \be\label{conventional}
Q_{BRST}=\oint [\;(\gamma^-\lambda_-)d_{--}+\bar\gamma \bar
G+vj+cT+c^{\kappa}T_{\kappa}+O(3)]\,, \ee where the first term in
(\ref{conventional}) is just the Berkovits BRST charge with
$\lambda=\gamma^-\lambda_-$, $T_{\kappa}$ are $\kappa$--symmetry
generators with $c^{\kappa}$ being the corresponding ghosts, and
$O(3)$ stands for cubic (anti)ghost terms.

It is known that, for example, in the case of the Berkovits--Vafa
embedding of a bosonic string into an $n=1$ fermionic string, and
an $n=1$ fermionic string into an $n=2$ fermionic string
\cite{twist}, there is a similarity transformation involving the
BRST charges of these theories \cite{similarity} (see also
\cite{higherN} for higher $n$), namely \be\label{R}
e^{-R}\,Q_{n+1}\,e^R=Q_n+Q_{top}\,, \ee where $Q_{top}$ is the
BRST charge of a topological sector of trivial cohomology.

It would be of interest to understand whether an analogous
transformation exists that relates the BRST charge
(\ref{conventional}) with the Berkovits charge (\ref{QB}). For
studying this point it might be useful to implement the
observation of McArthur \cite{ian} that the transformation
(\ref{R}) is related to a non--linear realization of the
symmetry associated with $Q_{n+1}$.

\section{Generalization to a supergravity--SYM background}
\setcounter{equation}0
 Using results of \cite{odatonin}, we can generalize the above
 consideration to the superembedding action
(\ref{seaction}) which describes the heterotic string propagating
in a curved background which satisfies the constraints of $N=1$,
$D=10$ supergravity interacting with $N=1$, $D=10$
super--Yang--Mills. The detailed analysis of the $N=1$, $D=10$
supergravity--SYM constraints relevant to the consideration below
may be found in \cite{berkhowe}. Upon complexification and the
gauge fixing of the pullback of the spinor supervielbein $\bar
E^{\underline\alpha}_-$ (\ref{E}), which replaces $\bar D_-\Theta$
and $\bar\lambda_-$ of the flat case, we get the generalization of
the component action (\ref{compact}) to the curved superbackground
in the following form \bea\label{compactiB} I_{SE}&=&\half\int
d^2\xi\left[E^{\underline a}_{++}E_{--\underline a}\,+ \,
B_{++,--}+ \bar\psi_+(D_{--}+ A_{--})\psi_+
+\psi_+(D_{--} + A_{--})\bar\psi_+\right]\nonumber\\
&+&\half\int d^2\xi\, p_{++\underline a}\,(Y^-\G^{\underline
a}\G_{\underline b}\lambda_-\,E_{--}^{\underline b})\,,
\eea where now
$\lambda^{\underline\alpha}_-=E^{\underline\alpha}_-$, and
$E^{\underline A}_{\pm\pm}$, $E^{\underline\alpha}_-$ and
$B_{++,--}$ stand for the leading $(\eta=\bar\eta=0)$ components
of the pullbacks of the supervielbein (\ref{E}) and of the NS--NS
two--form $B^{(2)}$. $A_{--}^{IJ}=\partial_{--}Z^{\underline
M}A_{\underline M}^{IJ}$ is the pullback of the super--Yang--Mills
potential.

The variations of $\lambda^{\underline\alpha}_-$, the Lagrange
multiplier $p_{++\underline a}$ and the heterotic fermions
$\psi_+^I$ under the $n=(0,2)$ supersymmetry and
$\kappa$--symmetry transformations (\ref{hatk}) are now
\be\label{la} \delta\lambda^{\underline\alpha}_-=\delta
Z^{\underline M}\Omega_{\underline
M\underline\beta}^{~~~\underline\alpha}\lambda^{\underline\beta}_-\,,
\ee \be\label{now} \delta p_{++\underline a}
=(E_{++}+W^{IJ}\bar\psi_+^I\psi_+^J)\,\Gamma_{\underline a}\delta
Z^{\underline M}E_{\underline M}\,-\half\,p_{++\underline b}\delta
Y\;\Gamma^{\underline b}\Gamma_{\underline a}\lambda\,, \ee
\be\label{psi} \delta \psi_+^I = -\delta Z^{\underline
M}E_{\underline M}^{\underline\alpha}
A^{IJ}_{\underline\alpha}\psi^J_+\,, \ee where
$\Omega_{++\underline\alpha}^{~~~\underline\beta}=dZ^{\underline
M}\Omega_{\underline M\underline\alpha}^{~~~\underline\beta}=
d\Phi\delta_{\underline\alpha}^{\underline\beta}+\Omega^{\underline
a\underline
b}\Gamma_{\underline{ab}\underline\alpha}^{~~~\underline\beta}$ is
a spin connection containing the differential of the dilaton
superfield $\Phi(Z)$ and the $SO(1,9)$ spin connection,
$W^{{\underline\alpha}IJ}=\Gamma^{\underline
a\underline\alpha\underline\beta}F^{IJ}_{\underline
a\underline\beta}$ and $F^{IJ}=(dA+A\wedge A)^{IJ}$ is a
constrained super--Yang--Mills stress tensor.

Note that the $n=(0,2)$ SUSY variation (\ref{la}) of
$\lambda^{\underline\alpha}_-$ induces local $SO(1,9)$ Lorentz rotations with
the parameter $\delta Z^{\underline M}\Omega_{\underline M}^{\underline
a\underline b}\Gamma_{\underline{ab}}$. This Lorentz transformation could be
compensated in (\ref{compactiB}) by the corresponding Lorentz rotations of
$p_{++\underline a}$, $E^{\underline b}_{--}$ and $Y^-_{\underline\alpha}$. But
by construction (\ref{Y}) $Y^-_{\underline\alpha}$ is not Lorentz covariant,
the `anomalous' Lorentz variation of $Y^-_{\underline\alpha}$ being
\be\label{anoY} \delta
Y_{\underline\alpha}=\delta Z^{\underline M}(\Omega_{\underline
M\underline\alpha}^{~~~\underline\beta}\,Y_{\underline\beta}-
Y_{\underline\alpha}\; \Omega_{\underline
M\underline\gamma}^{~~~\underline\beta}\;Y_{\underline\beta}\;\lambda^
{\underline\gamma})\,, \quad \lambda \delta Y \equiv 0\,.
\ee
In order to cancel this variation of $Y_{\underline\alpha}$ the
variation of the Lagrange multiplier $p_{++\underline a}$ should
acquire the last term (\ref{now}).

We now twist and gauge fix the model as in Section 5 with the
gauge fermion \be\label{fermiB} {\cal F}=p_{++\underline
a}\,d_{--}\Gamma^{\underline a}Y+\omega_{--}(1-{\cal
P})(E_{++}+W^{IJ}\bar\psi_+^I\psi_+^J)+\half p_{++\underline
a}\,\omega_{--}(1-{\cal P})\Gamma^{\underline a}\{Q_B,Y\}\,, \ee
the resulting action $I_{SE}+\int d^2\xi \{Q_B,{\cal F}\}$ is the
Berkovits action in the $N=1$, $D=10$ supergravity and
super--Yang--Mills background \cite{odatonin,berkhowe}
\bea\label{BB} I_B&=&\half\, \int d^2\xi\left [ E_{++}^{\underline
a}E_{--\underline a}+ B_{++,--}+ \bar\psi_+(\partial_{--}+
A_{--})\psi_+
+\psi_+(\partial_{--}+ A_{--})\bar\psi_+\right]\nonumber\\
&& +\,\int
d^2\xi\left[d_{--\underline\alpha}(E^{\underline\alpha}_{++}
+W^{{\underline\alpha}IJ}\bar\psi_+^I\psi_+^J)
+\omega_{--}(\partial_{++}+\Omega_{++}+ \half
U^{IJ}\bar\psi_+^I\psi_+^J)\lambda \right]\,, \eea where
$\Omega_{++\underline\alpha}^{~~~\underline\beta}=\partial_{++}
Z^{\underline M}\Omega_{\underline
M\underline\alpha}^{~~~\underline\beta}$ is a spin connection,
$U^{IJ\underline\beta}_{\underline\alpha}=\nabla_{\underline\alpha}
W^{IJ\underline\beta}$ and
$\nabla_{\underline\alpha}=E_{\underline\alpha}^{\underline
M}[\partial_{\underline M}+\Omega_{\underline M}+A_{\underline
M}]$ is a target superspace covariant spinor derivative (see
\cite{berkhowe} for details). Note that in (\ref{BB}) we used the
following BRST relations \be
\{Q_B,Y_{\underline\alpha}\}=\lambda^{\underline\delta}\,
(\Omega_{\underline
\delta\underline\alpha}^{~~~\underline\beta}\,Y_{\underline\beta}
-Y_{\underline\alpha}\;
\Omega_{\underline\delta\underline\gamma}^{~~~\underline\beta}\;
Y_{\underline\beta}\; \lambda^{\underline\gamma})\,, \ee \be
\{Q_B,p_{++\underline\a}\}=-(E_{++}+W^{IJ}\bar\psi_+^I\psi_+^J)
\,\Gamma_{\underline a}\lambda-\half p_{++\underline
b}\{Q_B,Y\}\,\Gamma^{\underline b}\Gamma_{\underline a}\lambda \,,
\ee \be \{Q_B,d_{--}\}=-\lambda\Gamma_{\underline a}E^{\underline
a}_{--}\,, \ee \be\{Q_B,E_{++}^{\underline\alpha}\}=
[(\partial_{++}+\Omega_{++})\lambda]^{\underline\alpha}\,, \quad
\{Q_B,W^{IJ}\bar\psi_+^I\psi_+^J\}=\half \lambda
U^{IJ}\bar\psi_+^I\psi_+^J\,. \ee

This concludes the reconstruction of the link between the
superembedding and the pure spinor formulation of the superstring.

\section{Conclusion}
We have obtained the pure spinor BRST charge, the antighost, and
the corresponding action for the heterotic string introduced by
Berkovits by gauge--fixing and twisting the $n=(0,2)$, $d=2$
superdiffeomorphism invariant heterotic string action of the
geometrical superembedding formulation. Since the superembedding
is classically equivalent to the Green--Schwarz formulation, we
have thus related , via superembedding, the Green--Schwarz and the
pure spinor superstring action.

As a natural generalization of these results one may wonder how to
demonstrate the analogous relation for the type II $D=10$
superstrings and the $D=11$ supermembrane. To this end one should
know the $n=2$ superdiffeomorphism invariant form of
superembedding actions for these objects. By now only an
$n=(1,1)$, $d=2$ worldsheet superfield action for a type IIB
superstring \cite{gal2} and an $n=1$, $d=3$ worldvolume superfield
action for a supermembrane \cite{pst} (see also \cite{hrs}) have
been constructed. The $n=(1,1)$, $d=2$ formulation of the type IIA
superstring can be obtained from the supermembrane action by the
double--dimensional reduction. The problem is to promote these
actions to be $n=(2,2)$, $d=2$ and $n=2$, $d=3$ supersymmetric.

The Berkovits quantization method has given a recipe of how to compute, in a
manifestly super--Poincar\'e covariant manner, tree--level amplitudes of
quantum superstring states. This is useful for the analysis of the quantum
superstring theory in super--Yang--Mills and supergravity backgrounds,
especially with nonzero Ramond--Ramond fields, and for the derivation of the
corresponding effective field theories. An important problem for applications
of the pure spinor quantization method is to understand a systematic way of
constructing one-- and higher--loop amplitudes. As we have already mentioned
this should involve the composite antighost field (\ref{b}) and requires an
additional insight into the proof of Lorentz--invariance of the higher--loop
amplitudes.

\section*{Acknowledgements}
It is a pleasure to thank N. Berkovits for interest in this work and for
illuminating discussions on his formulation of superstrings. We are also
grateful to I. Bandos, P. Budinich, F. Fucito, P.~A.~Grassi, P. Pasti, I.
Pesando, G. Policastro and K. Skenderis for useful discussions. This work was
partially supported by the European Community's Human Potential Programme under
contract HPRN-CT-2000-00131 Quantum Spacetime (L.M., M.M., D.S. and M.T.), by
the Grant N 383 of the Ukrainian State Fund for Fundamental Research (D.S.) by
the INTAS Research Project N 2000-254 (M.M., D.S., and M.T.) and by the Grant N
14540277 from the Japan Society for the Promotion of Science (I.O.).


\begin{thebibliography}{99}

\bibitem{report}
D. Sorokin, {\em Superbranes and Superembeddings}, {\PRP \ {\bf
329} (2000) 1}, [arXiv:hep--th/9906142];
 {\em Introduction to the Superembedding Description of
Superbranes}, [arXiv:hep--th/0105102].

\bibitem{siegel}
J.~A.~de Azcarraga and J.~Lukierski, {\em Supersymmetric Particles
with Internal Symmetries and Central Charges}, Phys.\ Lett.\
{\bf B113} (1982) 170;
{\em Supersymmetric Particles in
N=2 Superspace: Phase Space Variables and Hamiltonian Dynamics},
Phys.\ Rev.\ {\bf D 28} (1983) 1337.\\
W. Siegel, {\em Hidden Local Supersymmetry in the Supersymmetric
Particle Action}, {\PL \ {\bf B128} (1983) 397.}

\bibitem{stv}
D.~P.~Sorokin, V.~I.~Tkach and D.~V.~Volkov, {\em Superparticles,
Twistors and Siegel Symmetry}, Mod.\ Phys.\ Lett.\ {\bf A4} (1989)
901.
\\
D.~P.~Sorokin, V.~I.~Tkach, D.~V.~Volkov and A.~A.~Zheltukhin,
{\em From the Superparticle Siegel Symmetry to the Spinning
Particle Proper Time Supersymmetry}, Phys.\ Lett.\ {\bf B216}
(1989) 302.

\bibitem{gal}
A.~Galperin and E.~Sokatchev, {\em A Twistor Like D = 10
Superparticle Action with Manifest N=8 Worldline Supersymmetry},
Phys.\ Rev.\ {\bf D 46} (1992) 714, [arXiv:hep--th/9203051].

\bibitem{berk}
N.~Berkovits, {\em A Covariant Action for the Heterotic
Superstring with Manifest Space--Time Supersymmetry and World
Sheet Superconformal Invariance}, Phys.\ Lett.\ {\bf B232} (1989)
184.

\bibitem{mario}
M.~Tonin, {\em World Sheet Supersymmetric Formulations of
Green--Schwarz Superstrings}, Phys.\ Lett.\ {\bf B266} (1991) 312;
 {\em Kappa Symmetry as World Sheet Supersymmetry in D = 10
Heterotic Superstring}, Int.\ J.\ Mod.\ Phys.\ {\bf A7} (1992)
6013.

\bibitem{berk8}
 N.~Berkovits, {\em Twistors, N=8 Superconformal Invariance and the
Green--Schwarz Superstring}, Nucl.\ Phys.\ {\bf B358} (1991) 169.

\bibitem{berktwistor}
N. Berkovits, {\em The Heterotic Green--Schwarz Superstring on an
$N=(2,0)$ Superworldsheet}, {\NP \ {\bf B379} (1992) 96},
[arXiv:hep--th/9201004].

\bibitem{gal1}
F.~Delduc, E.~Ivanov and E.~Sokatchev, {\em Twistor Like Superstrings with D =
3, D = 4, D = 6 Target Superspace and N=(1,0), N=(2,0), N=(4,0) World Sheet
Supersymmetry}, Nucl.\ Phys.\ B {\bf 384} (1992) 334 [arXiv:hep--th/9204071].\\
F.~Delduc, A.~Galperin, P.~S.~Howe and E.~Sokatchev, {\em A
Twistor Formulation of the Heterotic D = 10 Superstring with
Manifest (8,0) World Sheet Supersymmetry}, Phys.\ Rev.\ {\bf D 47}
(1993) 578, [arXiv:hep--th/9207050].

\bibitem{igo}
I.~Bandos, {\em Superembedding Approach and S-Duality: A Unified Description of
Superstring and Super--D1--Brane}, Nucl.\ Phys.\ B {\bf 599} (2001) 197
[arXiv:hep--th/0008249].

\bibitem{pt}
P.~Pasti and M.~Tonin, {\em Twistor Like Formulation of the
Supermembrane in D = 11}, Nucl.\ Phys.\ {\bf B418} (1994) 337,
[arXiv:hep--th/9303156].\\
E.~Bergshoeff and E.~Sezgin, {\em Twistor--like Formulation of
Super p-Branes}, Nucl.\ Phys.\ {\bf B422} (1994) 329,
[arXiv:hep--th/9312168].

\bibitem{bpstv}
I.~A.~Bandos, D.~P.~Sorokin, M.~Tonin, P.~Pasti and D.~V.~Volkov,
{\em Superstrings and Supermembranes in the Doubly Supersymmetric
Geometrical Approach}, Nucl.\ Phys.\ {\bf B446} (1995) 79,
[arXiv:hep--th/9501113].


\bibitem{hs}
P.~S.~Howe and E.~Sezgin, {\em Superbranes}, Phys.\ Lett.\ {\bf
B390} (1997) 133, [arXiv:hep--th/9607227];
{\em D = 11, p = 5}, Phys.\ Lett.\ {\bf B394} (1997) 62,
[arXiv:hep--th/9611008].

\bibitem{ss}
S.~J.~Gates and H.~Nishino, {\em D = 2 Superfield Supergravity, Local
(Supersymmetry)**2 and Nonlinear Sigma Models}, Class.\ Quant.\
Grav.\ {\bf 3} (1986) 391.\\
R. Brooks, F. Muhammed and S. J. Gates Jr., {\em Matter Coupled to
D = 2 Simple Unidexterous Supergravity, Local (Supersymmetry)**2
and Strings}, {Class. Quantum Grav.} {\bf 3} (1986) 745.\\
J. Kowalski--Glikman, {\em Doubly Graded Sigma Model with
Torsion},
{Phys. Lett. \/} {\bf B180} (1986) 358.\\
 J.~Kowalski--Glikman, J.~W.~van Holten, S.~Aoyama and
J.~Lukierski, {\em The Spinning Superparticle}, Phys.\ Lett.\ {\bf
B201} (1988) 487.\\
A. Kavalov and R.L. Mkrtchyan, {\em Spinning Superparticles},
Preprint
Yer.PhI 1068(31)--88, Yerevan, 1988 (unpublished).\\
J.~M.~Fisch, {\em The N=1 Spinning Superstring: A Model of the
Super Superstring}, Phys.\ Lett.\ {\bf B219} (1989) 71.

\bibitem{cartan}
E. Cartan, {\em Lecons sur la Theorie des Spineurs}, Hermann,
Paris, 1937.\\
C. Chevalley, {\em The Algebraic Theory of Spinors}, Columbia
U.P., New York, 1954.

\bibitem{budinich}
P. Budinich, {\em From the Geometry of Pure Spinors with their
Division Algebras to Fermion's Physics}, [arXiv:hep--th/0107158].
\\
P. Budinich and A. Trautman, {\em Fock Space Description of Simple
Spinors}, J. \ Math. \ Phys. {\bf 30} (1989) 2125.

\bibitem{tonintwistor}
M. Tonin, {\em Twistor--like Formulation of Heterotic Strings}, At
{X Italian conference on general relativity and gravitational
physics -- Bardonecchia, September $1-5$ (1992), in ``Bardonecchia
1992, General relativity and gravitational physics'', 435-452,
[arXiv:hep--th/9301055].}


\bibitem{nilsson}
B.E.W. Nilsson, {\em Pure Spinors as Auxiliary Fields in the
Ten--dimensional Supersymmetric Yang--Mills Theory}, Class. \
Quant. \ Grav. {\bf 3} (1986) L41.\\

\bibitem{Howelines}
P. Howe, {\em Pure Spinors Lines in Superspace and
Ten--dimensional Supersymmetric Theories}, Phys. \ Lett. {\bf
B258} (1991) 141; {\em Pure Spinors, Function Superspaces and
Supergravity Theories in Ten--Dimensions and Eleven--Dimensions},
Phys. \ Lett. \ {\bf B273} (1991) 90.

\bibitem{berkassault}
N. Berkovits, {\em Covariant Quantization of the Green--Schwarz
Superstring in a Calabi--Yau Background}, {\NP \, {\bf B431}
(1994) 258}, [arXiv:hep--th/9404162];
{\em A New Description of the Superstring}, {in
Proceedings to VIII Jorge Swieca Summer School on particles and
fields, p.490, World Scientific Publishing, 1996, [arXiv:hep--th/9604123];}
{\em Quantization of the Superstring with Manifest
$U(5)$ Super--Poincar\'e Covariance}, {\PL \, {\bf B457} (1999)
94}, [arXiv:hep--th/9902099].

\bibitem{nis}
E.~R.~Nissimov and S.~J.~Pacheva, {\em Manifestly Superpoincar\'e
Covariant Quantization of the Green--Schwarz Superstring}, Phys.\
Lett.\ {\bf B202} (1988) 325.
\\
E.~Nissimov, S.~Pacheva and S.~Solomon, {\em Covariant Canonical
Quantization of the Green--Schwarz Superstring}, Nucl.\ Phys.\
{\bf B297} (1988) 349.

\bibitem{rakh}
R.~Kallosh and M.~Rakhmanov, {\em Covariant Quantization of the
Green--Schwarz Superstring}, Phys.\ Lett.\ {\bf B209} (1988) 233.

\bibitem{ig}
I.~A.~Bandos and A.~A.~Zheltukhin, {\em Spinor Cartan Moving N
Hedron, Lorentz Harmonic Formulations of Superstrings, and Kappa
Symmetry}, JETP Lett.\ {\bf 54} (1991) 421 [Pisma Zh.\ Eksp.\
Teor.\ Fiz.\ {\bf 54} (1991) 421];
{\em Green--Schwarz Superstrings in Spinor Moving Frame
Formalism}, Phys.\ Lett.\ {\bf B288} (1992) 77.

\bibitem{berkpurespinor}
N. Berkovits, {\em Super--Poincar\'e Covariant Quantization of the
Superstring}, {\JHEP \ {\bf 0004} (2000) 018},
[arXiv:hep--th/0001035].
\\
N. Berkovits and B. Carlini Valillo, {\em Consistency of
Super--Poincar\'e Covariant Superstring Tree Amplitudes}, {\JHEP \
{\bf 0007} (2000) 015}, [arXiv:hep--th/0004171].
\\
N. Berkovits, {\em Covariant Quantization of the Superstring},
{\IJMP\, {\bf A16} (2001) 801}, [arXiv:hep--th/0008145].
\\
N. Berkovits and O. Chandia, {\em Lorentz Invariance of the Pure
Spinor BRST Cohomology for the Superstring}, {\PL \ {\bf B514}
(2001) 394}, [arXiv:hep--th/0105149].

\bibitem{berkrelating}
N. Berkovits, {\em Relating the RNS and Pure Spinor Formalism for the
Superstring}, {\JHEP \ {\bf 0108} (2001) 026},
[arXiv:hep--th/0104247].

\bibitem{purespectrum}
N. Berkovits, {\em Cohomology in the Pure Spinor Formalism for the
Superstring}, {\JHEP \ {\bf 0009} (2000) 046},
[arXiv:hep--th/0006003].
\\
N. Berkovits and O. Chandia, {\em Massive Superstring Vertex Operator
in $D=10$ Superspace}, {[arXiv:hep--th/0204121].}
\\
G.~Trivedi, {\em Correlation Functions in Berkovits' Pure Spinor Formulation},
[arXiv:hep--th/0205217].


\bibitem{berkother}
N. Berkovits, {\em Covariant Quantization of the Superparticle Using
Pure Spinors}, {\JHEP\, {\bf 0109} (2001) 016},
[arXiv:hep--th/0105050];
{\em Towards Covariant Quantization of the
Supermembrane}, {[arXiv:hep--th/0201151]}.

\bibitem{pershin}
N.~Berkovits and V.~Pershin, {\em Supersymmetric Born--Infeld from
the pure spinor formalism of the open superstring},
hep--th/0205154.

\bibitem{odatonin}
I. Oda and M. Tonin, {\em On the Berkovits Covariant Quantization of
GS Superstring}, {\PL \ {\bf B520} (2001) 398},
[arXiv:hep--th/0109051].

\bibitem{twist}
E. Witten, {\em Topological Quantum Field Theory}, {\CMP \ {\bf 117}
(1988) 355;}
\\
N. Berkovits and C. Vafa, {\em On the Uniqueness of String
Theory},
{\MPL \ {\bf A9} (1994) 653, [arXiv:hep--th/9310170];}
{\em $N=4$ Topological Strings}, {\NP \
{\bf B433} (1995) 123}, [arXiv:hep--th/9407190].
\\
N. Berkovits, C. Vafa and E. Witten, {\em Conformal Field Theories of
AdS Background with Ramond--Ramond Flux}, {\JHEP \ {\bf 9903}
(1999) 018}, [arXiv:hep--th/9902098].

\bibitem{igor}
I.~Bandos and T.~Bandos, {\em Lorentz Harmonics and Superfield
Action: D = 10, N = 1 Superstring}, Class.\ Quant.\ Grav.\ {\bf
18} (2001) 1907 [arXiv:hep--th/0010044].

\bibitem{senza}
P.~A.~Grassi, G.~Policastro, M.~Porrati and P.~van Nieuwenhuizen, {\em
Covariant Quantization of Superstrings without Pure Spinor Constraints},
[arXiv:hep--th/0112162].
\\
P.~A.~Grassi, G.~Policastro and P.~van Nieuwenhuizen, {\em The
Massless Spectrum of Covariant Superstrings}, [arXiv:hep--th/0202123].

\bibitem{alaSiegel}
W. Siegel, {\em Classical Superstring Mechanics}, {\NP \ {\bf B263}
(1986) 93.}

\bibitem{similarity}
H.~Ishikawa and M.~Kato, {\em Note on N=0 String as N=1 String},
Mod.\ Phys.\ Lett.\ {\bf A9} (1994) 725, [arXiv:hep--th/9311139].\\
F.~Bastianelli,
{\em A Locally Supersymmetric Action for the Bosonic String},
Phys.\ Lett.\ {\bf B322} (1994) 340, [arXiv:hep--th/9311157].\\
N.~Ohta and J.~L.~Petersen, {\em N=1 from N=2 Superstrings}, Phys.\
Lett.\ {\bf B325} (1994) 67, [arXiv:hep--th/9312187].\\
M.~Kato, {\em Physical Spectra in String Theories: BRST Operators and
Similarity Transformations}, [arXiv:hep--th/9512201].

\bibitem{higherN} F.~Bastianelli, N.~Ohta and J.~L.~Petersen,
{\em A Hierarchy of Superstrings}, Phys.\ Rev.\ Lett.\ {\bf 73}
(1994) 1199, [arXiv:hep--th/9403150].

\bibitem{ian}
I.~N.~McArthur, {\em Gauging of Nonlinearly Realized Symmetries},
Nucl.\ Phys.\ {\bf B452} (1995) 456, [arXiv:hep--th/9504160].

\bibitem{berkhowe}
N. Berkovits and P. Howe, {\em Ten--Dimensional Supergravity
Constraints from the Pure Spinor Formalism for the Superstring},
{[arXiv:hep--th/0112160].}

\bibitem{gal2}
A.~Galperin and E.~Sokatchev, {\em A Twistor Formulation of the
Nonheterotic Superstring with Manifest World Sheet Supersymmetry},
Phys.\ Rev.\ {\bf D 48} (1993) 4810, [arXiv:hep--th/9304046].

\bibitem{pst}
P.~Pasti, D.~P.~Sorokin and M.~Tonin, {\em Superembeddings,
Partial Supersymmetry Breaking and Superbranes}, Nucl.\ Phys.\
{\bf B591} (2000) 109, [arXiv:hep--th/0007048].

\bibitem{hrs}
P.~S.~Howe, O.~Raetzel and E.~Sezgin, {\em On Brane Actions and
Superembeddings}, JHEP {\bf 9808} (1998) 011,
[arXiv:hep--th/9804051].
\\
J.~M.~Drummond and P.~S.~Howe, {\em Codimension Zero
Superembeddings}, Class.\ Quant.\ Grav.\ {\bf 18} (2001) 4477,
[arXiv:hep--th/0103191].
\\
P.~S.~Howe and U.~Lindstrom, {\em Kappa--Symmetric Higher
Derivative Terms in Brane Actions}, Class.\ Quant.\ Grav.\ {\bf
19} (2002) 2813, [arXiv:hep--th/0111036].

\end{thebibliography}
\end{document}